\documentclass[12pt]{article}  
\usepackage{todonotes}
\usepackage[authoryear,round]{natbib}
\usepackage{graphicx}
\usepackage{latexsym}
\usepackage[left=1.5in,top=1.5in,right=1.5in,bottom=1.5in]{geometry}\usepackage{amsmath}
\usepackage{booktabs}
\usepackage{color}
\usepackage{tikz}
\usepackage{algorithmic}
\usetikzlibrary{shapes}
\usetikzlibrary{arrows}
\definecolor{darkblue}{rgb}{0,0.4,0.9}
\definecolor{gray10}{rgb}{0.1,0.1,0.1}
\definecolor{gray20}{rgb}{0.2,0.2,0.2}
\definecolor{gray30}{rgb}{0.3,0.3,0.3}
\definecolor{gray40}{rgb}{0.4,0.4,0.4}
\definecolor{gray60}{rgb}{0.6,0.6,0.6}
\definecolor{gray80}{rgb}{0.8,0.8,0.8}
\definecolor{gray90}{rgb}{0.9,0.9,.9}
\definecolor{gray95}{rgb}{0.95,0.95,.95}
\definecolor{gray96}{rgb}{0.96,0.96,.96}
\definecolor{lgreen} {RGB}{180,210,100}
\definecolor{dblue}  {RGB}{20,66,129}
\definecolor{ddblue} {RGB}{11,36,69}
\definecolor{lred}   {RGB}{220,0,0}
\definecolor{nred}   {RGB}{224,0,0}
\definecolor{norange}{RGB}{230,120,20}
\definecolor{nyellow}{RGB}{255,221,0}
\definecolor{ngreen} {RGB}{98,158,31}
\definecolor{dgreen} {RGB}{78,138,21}
\definecolor{nblue}  {RGB}{28,130,185}
\definecolor{jblue}  {RGB}{20,50,100}
\definecolor{nnyellow}{RGB}{235,200,0}
\definecolor{purple}{RGB}{150, 0, 120}
\definecolor{sgGreen} {RGB}{20, 180, 50}
\definecolor{revised}{rgb}{0,0,0.9}

\newtheorem{theorem}{Theorem}

\newcommand{\openr}{\hbox{${\rm I\kern-.2em R}$}}
\newcommand{\openn}{\hbox{${\rm I\kern-.2em N}$}}

\newcommand{\subfamily}[1]{\ensuremath\mathcal{G}_{#1}}		
\newcommand{\familysize}[1]{\ensuremath{|\subfamily#1|}}		
\newcommand{\minp}[1]{\ensuremath{p^{\mathrm{min}}_{#1}}}
\newcommand{\minh}[1]{\ensuremath{h^{\mathrm{min}}_{#1}}}		

\makeatletter
\renewcommand{\@todonotes@textwidth}{3cm}
\makeatother

\title{Controlling Familywise Error When Rejecting at Most One Null Hypothesis Each From a Sequence of Sub-Families of Null Hypotheses}

\author{Geoffrey I.\ Webb\\
	Monash University\\
	{\tt geoff.webb@monash.edu}\\
	Mark van der Laan\\
	University of California, Berkeley\\
	{\tt laan@berkeley.edu}} \date{\today}

\begin{document}
\maketitle
\begin{abstract}
We present a procedure for controlling FWER when sequentially considering successive subfamilies of null hypotheses and rejecting at most one from each subfamily. Our procedure differs from previous procedures for controlling FWER by adjusting the critical values that are applied in subsequent rejection decisions by subtracting from the global significance level $\alpha$ quantities based on the p-values of rejected null hypotheses and the numbers of null hypotheses considered.
\end{abstract}


\section{Introduction}
We present a procedure for strictly controlling the Familywise Error Rate when rejecting a single null hypothesis from each subfamily in a sequence of subfamilies of null hypotheses, where each rejection decision is made without knowledge of subsequent subfamilies. 

Our procedure is a more powerful variant of a procedure presented by \cite{WebbPetitjean16}.  These procedures differ in form from previous multiple testing procedures by adjusting the critical value applied to subsequent subfamilies based on the observed values of test statistics for null hypotheses in prior subfamilies. 

We identify the assumptions of the procedure, use Monte Carlo simulations to elucidate properties of the procedure under differing scenarios when the assumptions are satisfied, and provide analytical and Monte Carlo simulation results to demonstrate scenarios under which FWER is not controlled when the assumptions are violated.

\subsection{Set-up}
Let ${\bf O}_n$ be a random variable with probability distribution $P_0^n$. Suppose we observe a realization of this random variable representing our observed data. 
Let ${\cal G}_t$, $t=1,\ldots,T$, be an ordered sequence of subfamilies of null hypotheses, where ${\cal G}_t=\{H_0(t,j):j=1,\ldots,m_t\}$  consists of $m_t$ null hypotheses about the data distribution $P_0^n$. Let $T_n(t,j)$ be a test-statistic for null hypothesis $H_0(t,j)$, $j=1,\ldots,m_t$, $t=1,\ldots,T$.
Let ${\cal N}^t=\{j: H_0(t,j)\mbox{ is true}\}$ be the set of true null hypotheses in ${\cal G}_t$, and let ${\cal F}^t=\{j:H_0(t,j)\mbox{ is false}\}$ be the set of false null hypotheses in ${\cal G}_t$. Let ${\cal N}=\{(t,j):H_0(t,j)\mbox{ is true}\}$ and ${\cal F}=\{(t,j):H_0(t,j)\mbox{ is false}\}$ be the sets of true and false null hypotheses among all null hypotheses.

{\bf P-values:}  Let $P_n(t,j)$ be a p-value implied by $T_n(t,j)$. It is assumed that if $H_0(t,j)$ is true, then $\mbox{Pr}(P_n(t,j)\leq \alpha)\leq \alpha$ for all $\alpha\in [0,1]$. 
In  other words, the $p$-value, which is just a transformation of the test-statistic, satisfies its key property. For simplicity, we assume this to be true for the finite sample $O^{\cal N}$. As a result, our theorem establishes exact control of the family wise error, but in the often more realistic case that the null distributions of the test statistics are only known asymptotically so that $\lim_{n\rightarrow\infty}\mbox{Pr}(P_n(t,j)\leq \alpha)\leq \alpha$, our  results will provide asymptotic control of the family wise error.  

\sloppy Let ${\bf P}_n=(P_n(t,j):t,j)$ be the vector of $p$-values, and let ${\bf P}_n^{\cal N}=(P_n(t,j):(t,j)\in {\cal N})$ and ${\bf P}_n^{\cal F}=(P_n(t,j): (t,j)\in {\cal F})$ the vector of $p$-values for the true null hypotheses and false null hypotheses, respectively.
Let~$P_n(t)=\min_jP_n(t,j)$ be the minimum $p$-value for family ${\cal G}_t$ and let $J_n(t)=\arg\min_j P_n(t,j)$ identify  the null-hypothesis with the minimal $p$-value. 
Thus $P_n(t,J_n(t))=\min_jP_n(t,j)$. We also define $P_n^{\cal N}(t)=\min_{\{j: (t,j)\in {\cal N}\}}P_n(t,j)$ and $P_n^{\cal F}(t)=\min_{\{j:(t,j)\in {\cal F} \}}P_n(t,j)$ as the minimum of the p-values over the set of true and false null hypotheses in family ${\cal G}_t$, respectively. 
More precisely,
\begin{equation*}P_n^{\cal F}(t)=\begin{cases}\min_{\{j: (t,j)\in {\cal F}\}}P_n(t,j)&\mbox{if }{\cal F}^t\neq\emptyset\\1.0&\mbox{otherwise} \end{cases}
\end{equation*}
\begin{equation*}P_n^{\cal N}(t)=\begin{cases}\min_{\{j: (t,j)\in {\cal N}\}}P_n(t,j)&\mbox{if }{\cal N}^t\neq\emptyset\\1.0&\mbox{otherwise} \end{cases}
\end{equation*}

Our goal is to define a sequential multiple testing procedure that rejects at most one hypothesis per subfamily ${\cal G}_t$, making the decision as to whether or not to reject without knowledge of subsequent subfamilies and that controls the familywise error over all subfamilies ${\cal G}_t$, $t=1,\ldots,T$ at user supplied level $\alpha\leq1$.

\subsection{Sequential multiple testing procedure for a sequence of families of null hypotheses.}
We propose the following sequential multiple testing procedure that results in a set of rejections ${\cal R}=\{J_n(l):l=1,\ldots,K^*\}$.

\noindent{\bf Multiple Testing Procedure:}
\begin{algorithmic}[1]
	\STATE {\bf let} $K=1$.
	\STATE {\bf let} $\alpha(K)=\alpha$.
	\STATE {\bf let} ${\cal R}=\emptyset$.
	\STATE {\bf let} $K^*=0$.
	\WHILE{$K\leq T$ and $m_KP_n(K)\leq\alpha(K)$}
	\STATE {\bf let} ${\cal R}={\cal R}\cup (K,J_n(K))$.
	\STATE {\bf let} $\alpha(K+1)=\alpha(K)-(m_K-1)P_n(K)$.\label{line:adjust}
	\STATE {\bf let} $K=K+1$.
	\ENDWHILE
	\STATE {\bf let} $K^*=K-1$.
\end{algorithmic}

This procedure differs from that of \cite{WebbPetitjean16} at line \ref{line:adjust} where their procedure has instead {\bf let} $\alpha(K+1)=\alpha(K)-m_KP_n(K)$. By subtracting a smaller quantity from each successive $\alpha(K)$ our procedure is guaranteed to be uniformly more powerful.  Hence, our proof also provides a proof of correctness for this prior procedure.

\subsection{Theorem establishing family wise error control}
The following theorem proves that for each realization of the $p$-values ${\bf P}_n^{\cal F}$ of the false null hypotheses, the conditional probability of rejecting a true null hypothesis
is no greater than  
$\alpha$. Of course, this implies, in particular, that the marginal probability on any rejection of a true null is no greater than $\alpha$. 
The key assumption this theorem relies upon is that the $p$-values of the true nulls are independent of the $p$-values of the false nulls. 

\begin{theorem}\label{th1}
	Assume that  ${\bf P}_n^{\cal N}=(P_n(t,j):(t,j)\in {\cal N})$ is independent of ${\bf P}_n^{\cal F}=(P_n(t,j):(t,j)\in {\cal F})$.  Specifically, assume that for all possible realizations of ${\bf P}_n^{\cal F}$, $\mbox{Pr}(P_n(t,j)\leq \alpha)\leq \alpha$ for all $P_n(t,j):(t,j)\in {\cal N}$ and all $\alpha\in [0,1]$.
	Then, 
	\[
	\mbox{Pr}({\cal R}\cap {\cal N}=\emptyset\mid {\bf P}_n^{\cal F})\geq 1-\alpha .\]
\end{theorem}

{\bf Proof:}
In this proof we condition on ${\bf P}_n^{\cal F}$, so that all probabilities concern the random variable ${\bf P}_n^{\cal N}$.\newline
\noindent{\bf Scenario I:}
First, consider the scenario that
\begin{equation*}
\sum_{l=1}^{T-1}(m_l-1)P_n^{\cal F}(l)+m_TP_n^{\cal F}(T)\leq\alpha.
\end{equation*}
We note that this implies that  all the subfamilies contain at least one false null hypothesis.
The probability of a false rejection at the $i$-th subfamily, $i\in[1,T]$, is the probability that $P_n^{\cal N}(i)\leq P_n^{\cal F}(i)$, which is no greater than $(m_i-1)P_n^{\cal F}(i)$, where we use that $P_n^{\cal N}$ is a minimum over maximally $m_i-1$ true null hypotheses. 

The union from $i=1,\ldots,T$ represents the event that we have a false rejection.  This proves that the probability of a false rejection is no greater than $\sum_{l=1}^{T}(m_l-1)P_n^{\cal F}(l)<\sum_{l=1}^{T-1}(m_l-1)P_n^{\cal F}(l)+m_TP_n^{\cal F}(T)\leq\alpha$. 

\noindent{\bf Scenario II:}
The only alternative to Scenario I is the scenario that there exists a first  $j\in \{1,\ldots,T\}$ such that  $m_jP_n^{\cal F}(j)>\alpha-\sum_{l=1}^{j-1}(m_l-1)P_n^{\cal F}(l)$, and thus, for $i=1,\ldots,j{-}1$, we have $m_iP_n^{\cal F}(i)\leq\alpha-\sum_{l=1}^{i-1}(m_l-1)P_n^{\cal F}(l)$.
We note that this implies that the $i$-th subfamily has at least one false null hypothesis, $i=1,\ldots,j{-}1$, and that the probability of a false rejection of a true null hypothesis in ${\mathcal G}_1, \ldots, {\mathcal G}_{j-1}$ is no greater than $\sum_{l=1}^{j-1}(m_l-1)P_n^{\cal F}(l)$.

If there has been no false rejection in ${\mathcal G}_1, \ldots, {\mathcal G}_{j-1}$ this implies that for all $i=1,\ldots,j-1$, $P_n(i)=P_n^{\mathcal F}(i)$ as otherwise there would have been a false rejection of the true null hypothesis corresponding to $P_n(i)$.

In this scenario, the procedure rejects $J_n(j)$ if and only if $m_jP_n(j)\leq\alpha-\sum_{l=1}^{j-1}(m_l-1)P_n^{\cal F}(l)$ and hence the probability of a first false rejection at ${\mathcal G}_j\leq \alpha-\sum_{l=1}^{j-1}(m_l-1)P_n^{\cal F}(l)$.

The probability of the union of the two events of a rejection in ${\mathcal G}_1,\ldots {\mathcal G}_{j-1}$ and of a rejection in ${\mathcal G}_j$ but no rejection in ${\mathcal G}_1,\ldots {\mathcal G}_{j-1}$  is thus no greater than $\sum_{l=1}^{j-1}(m_l-1)P_n^{\cal F}(l)+\alpha-\sum_{l=1}^{j-1}(m_l-1)P_n^{\cal F}(l)=\alpha$.

$\Box$

\section{Discussion}

\subsection{Relationship to other approaches for controlling FWER}
The standard fixed sequence hypothesis test procedure \citep{maurer1995multiple,Hsu99}, where all of a fixed sequence of null hypotheses are tested at level $\alpha$, is a special case of our procedure where all subfamilies are of size 1 (all $m_i=1$).

Our procedure follows a fundamentally different strategy to gatekeeping procedures based on Bonferroni adjustments \citep{Bauer1998,Westfall2001,Dmitrienko2003,Chen2005}.
Gatekeeping procedures add the $\alpha(i)$ for rejected null hypotheses to the $\alpha(i)$ of subsequent hypotheses.  In contrast, our procedure subtracts from subsequent $\alpha(i)$ some portion of the previous $\alpha(i)$, based on the observed p-value of the rejected null hypotheses.

The approach also differs fundamentally from selective inference \citep{Taylor2015}.
First, our procedure controls FWER, while selective inference controls FDR.  Second, unlike our procedure, selective inference, does not use an explicit sequential order over subfamilies of null hypotheses.  Third, also unlike our procedure, selective inference rejects null hypotheses in order of ascending p-value until a function over the p-values of the null hypotheses exceed a threshold.

\subsection{Monte Carlo experiments}

To elucidate the statistical power of the technique, we conducted Monte Carlo simulations.  In all the following simulations we use $\alpha=0.05$.

In the first simulation we generated sets of null hypotheses, which were randomly assigned to be either true or false and were randomly assigned simulated p-values.  These simulations were governed by three parameters --- \emph{subfamilySize}: the size of each subfamily; \emph{pTrue}: the probability that a null hypothesis should be designated to be true; and \emph{maxFalsePVal}: the maximum simulated p-value to be assigned to a false null hypothesis.

The following procedure was used for this simulation.

\noindent{\bf Monte Carlo simulation procedure}
\begin{algorithmic}
	\STATE $\textit{flag}\leftarrow \textrm{false}$
	\STATE $\mathcal{R} \leftarrow \emptyset$
	\STATE $\textit{sumP}\leftarrow 0.0$
	\STATE $i\leftarrow 0$
	\WHILE{$\textit{flag}=\textrm{false}$}
	\STATE $i\leftarrow i+1$
	\STATE Generate $\subfamily{i}$
	\IF{$\textit{sumP}+\textit{\familysize{i}}\cdot \minp{i}\leq\alpha$}
	\STATE $\mathcal{R}\leftarrow \mathcal{R}\cup \{\minh{i}\}$
	\STATE $\textit{sumP}\leftarrow \textit{sumP}+\textit{\familysize{i}}\cdot \minp{i}$
	\ELSE
	\STATE $\textit{flag}\leftarrow \textrm{true}$
	\ENDIF
	\ENDWHILE
\end{algorithmic}

To generate each $\subfamily{i}$,  \emph{subfamilySize} simulated null hypotheses were generated. Each was designated as either true or false, with probability \emph{pTrue} of being designated true. Each true null hypothesis was assigned a simulated p-value drawn uniformly at random from $[0.0,1.0]$ and each false null hypothesis was assigned a simulated p-value drawn uniformly at random from $[0.0,0.1]$. Having lower p-values for false null hypotheses simulates the use of a test statistic that is useful for discriminating between true and false null hypotheses.

\emph{pTrue} was varied from $0.1$ to $1.0$ in steps of $0.1$ and \emph{subfamilySize} was set to each of the values $1$, $10$, $100$ and $1,000$,  creating a total of 40 treatments.  $1,000,000$ Monte Carlo simulations were conducted for each treatment and the FWER and average number of true discoveries per simulation determined.

\begin{figure}
	\includegraphics[width=\textwidth,trim=55pt 100pt 55pt 100pt]{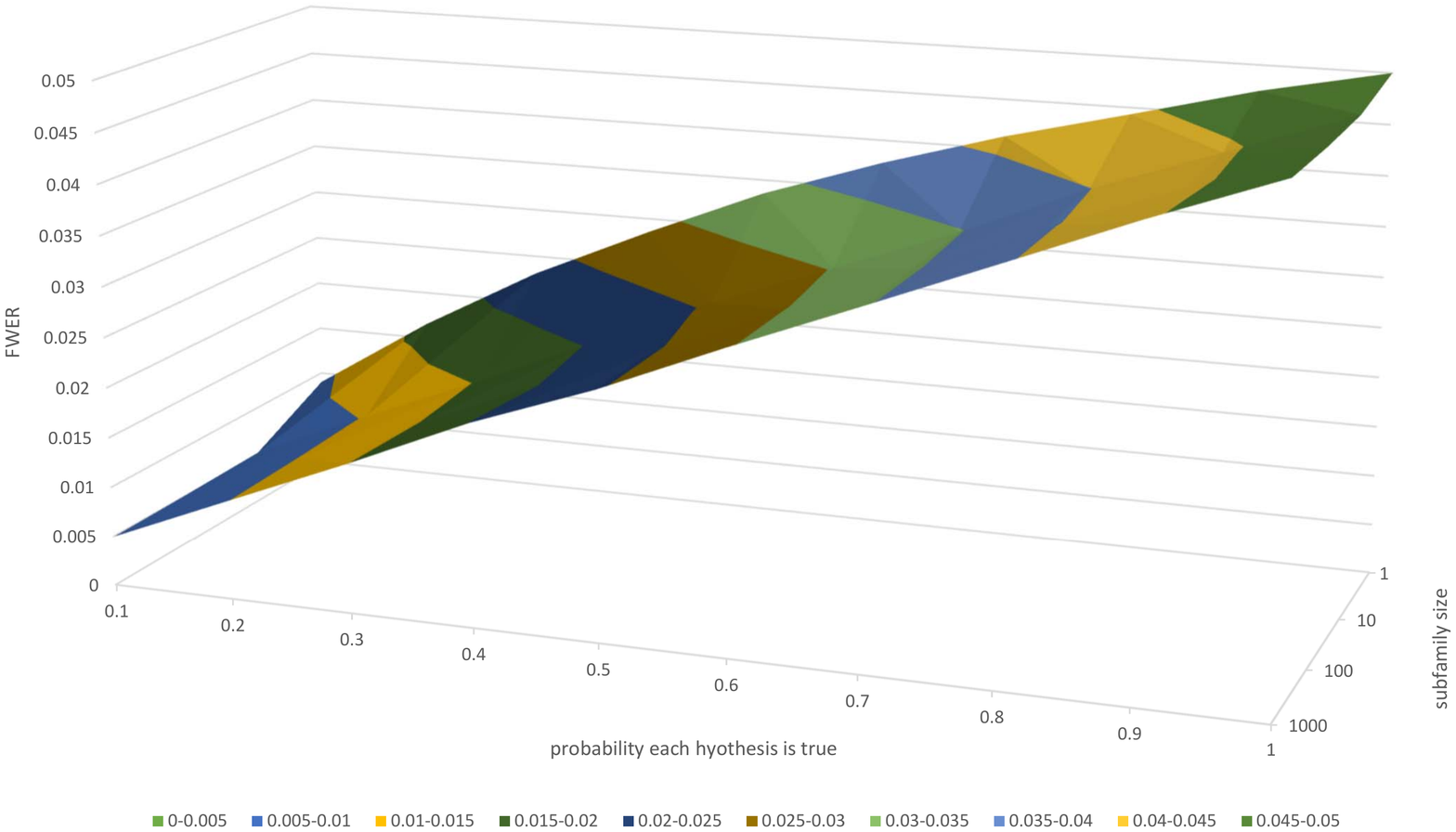}
	\caption{The FWER of SMT as the relative frequency of true to false null hypotheses is increased and the relative p-values of false relative to true null hypotheses is decreased}\label{fig:FWERBySubfamilySize}
\end{figure}%
Figure~\ref{fig:FWERBySubfamilySize} presents a surface chart showing the effect on FWER as the relative frequency of true to false null hypotheses is increased and as the subfamily size varies.  When pTrue is 1.0 and FWER is determined by whether a null hypothesis is rejected for the first subfamily or not, the probability of FWER is strictly controlled by the equivalent of a Bonferroni correction for the first subfamily. FWER falls as the proportion of false null hypotheses rises because the multiple test correction is allowing for the possibility that they are all true. 

Increasing subfamilysize also decreases FWER because the multple test correction allows for the worst case where the rejection regions of all null hypotheses are disjoint whereas in this simulation all null hypotheses are independent of one another. 

This simulation demonstrates the power of our procedure when its assumptions are satisfied, and show that it is most powerful when the ratio of false to true hypotheses is highest and subfamilysize is smallest.

We next demonstrate a scenario where violating the requirement that True and False null hypotheses be independent results in a failure to control FWER.

In this scenario we have one false null hypothesis, $A$ and two true null hypotheses, $B$ and $C$.
The experimental outcome on which $A$ and $B$ are based is the result of tossing an unbiased coin 17 times.  The experimental outcome on which $C$ is based is the result of tossing another coin 13 times. Both coins are unbiased, $Pr(heads)=0.5$. We choose 17 for the first experiment because it is the smallest number of tosses that has an outcome for a test for $Pr(heads) = 0.5$ that is close to 0.025, and 13 for the second because it is the smallest number of tosses that has an outcome for a test for $Pr(heads) = 0.5$ that is close to 0.05. 
$A = Pr(heads)\leq 0.1$,
$B = Pr(heads)\geq0.5$ and
$C = Pr(heads)\neq0.5$.

$\subfamily{1} = \{A, B\}$ and 
$\subfamily{2} = \{C\}$.

We proceed to $\subfamily{2}$ if either $A$ or $B$ is rejected. 

$A$ and $B$ are tested at $\alpha/2 = 0.025$.

There are 17 coin tosses and the rejection region for $B$ is 4 or fewer heads.  The probability of this outcome is 0.0245.

The rejection region for $A$ is 5 or more heads.

$Pr(17\ heads) = 7.6294E-06; Pr(16\ heads) = 0.0001; Pr(15\ heads) = 0.0010; Pr(14\ heads) = 0.0052; Pr(13\ heads) =  0.0182, \ldots$.
The respective p-values for $A$ are $1.00E-17$, $1.54E-15$, $1.117E-13$, $5.0689E-12$, $1.6122E-10$, $3.8152E-09$, $6.9586E-08$, $9.9978E-07$, $1.1464E-05$, $0.0001$, $0.0008$,  $0.00467$ and $0.0221$, meaning A will be rejected if there are 5 or more heads and the adjusted alpha for C will be respectively $0.05-1E-17$ to $0.05-0.0221$.  If $C$ were a maximally powerful true null hypothesis then the probability of it being rejected would be $7.6294E-06 \times (\alpha-1E-17) + 0.0001 \times (\alpha-1.54E-15)+ \ldots+ 0.0.47 x 0.0221 = 0.047$.  Adding this to the probability of false rejection of $B$ gives a FWER of 0.0715. 

However, as we are using coin tosses with a finite number of outcomes, $C$ is not maximally powerful. A Monte Carlo simulation of 1,000,000 repetitions of this scenario yielded a FWER of $0.0647$ demonstrating again that violation of the requirement that the true and false null hypotheses be independent of one another can lead to failure to control familywise error.

\section{Conclusion}

We have presented a novel procedure for controlling familywise error in a sequential testing scenario where at most one null hypothesis is to be rejected from each of a series of subfamilies of null hypotheses.  We have shown that this procedure requires only the assumption that the p-values for the true and false null hypotheses are independent of one another. This assumption is realistic in the context of stepwise model selection for which the procedure was developed. 

The procedure uses a novel mechanism of adjusting subsequent critical values by quantities based on the observed p-values of null hypotheses that are rejected. It remains a promising avenue for future research to investigate whether this strategy is more broadly applicable in other sequential testing scenarios.

\section*{Acknowledgments}
This research has been supported by the Australian Research Council under grant DP140100087.

\end{document}